\documentclass[9pt,journal,letterpaper,twocolumn]{IEEEtran}

\usepackage{cite}
\usepackage{epsfig}
\graphicspath{{/}}

\begin{document}

\title{Graphene Terahertz Plasmon Oscillators} 
\author{Farhan Rana \\% ,~\IEEEmembership{Member,~IEEE} \\
School of Electrical and Computer Engineering, \\ 
Cornell University, Ithaca, NY 14853}

%\thanks{Copyright (c) 2007 IEEE. Personal use of this material is permitted. However, permission to use this material for any other other purposes must be obtained from the IEEE by sending a request to pubs-permissions@ieee.org.}

\maketitle

\begin{abstract}
In this paper we propose and discuss coherent terahertz sources based on charge density wave (plasmon) amplification in two dimensional graphene. The coupling of the plasmons to interband electron-hole transitions in population inverted graphene layers can lead to plasmon amplification through stimulated emission. Plasmon gain values in graphene can be very large due to the small group velocity of the plasmons and the strong confinement of the plasmon field in the vicinity of the graphene layer. We present a transmission line model for plasmon propagation in graphene that includes plasmon dissipation and plasmon interband gain due to stimulated emission. Using this model, we discuss design for terahertz plasmon oscillators and derive the threshold condition for oscillation taking into account internal losses and also losses due to external coupling. The large gain values available at terahertz frequencies in graphene can lead to integrated oscillators that have dimensions in the 1-10 $\mu$m range.  
\end{abstract}   
 

%\IEEEpeerreviewmaketitle

\addtocounter{page}{-1}

\section{Introduction}
\begin{figure}[tb]
  \begin{center}
   \epsfig{file=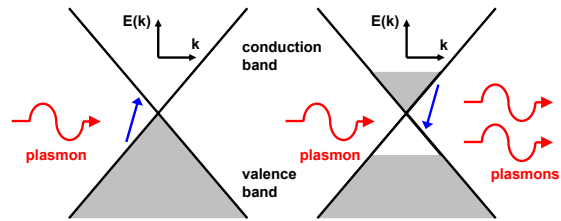,angle= 0,width=3.0 in}    
    \caption{(LEFT) Energy bands of graphene showing stimulated absorption of plasmons. (RIGHT) Population inverted graphene bands showing stimulated emission of plasmons.}
    \label{Fig1}
  \end{center}
\end{figure}

Graphene is a single two dimensional (2D) atomic layer of carbon atoms forming a dense honeycomb crystal lattice~\cite{dressel}. The electronic properties of graphene have generated tremendous interest in both experimental and theoretical arenas~\cite{chakra,ryzhii1,ryzhii2,nov1,nov2,zhang,heer,rana}. The energy dispersion relation of electrons and holes with zero (or close to zero) bandgap results in novel behavior of both single-particle and collective excitations~\cite{dressel,chakra,ryzhii1,ryzhii2,nov1,nov2,zhang,heer,rana}. The high mobility of electrons in graphene has prompted theoretical and experimental investigations into graphene based ultra high speed electronic devices such as field-effect transistors, pn-junction diodes, and terahertz devices~\cite{nov2,lundstrom,marcus,ryzhii1,ryzhii2,rana,ryzhii3,ryzhii4}. Negative conductance at terahertz frequencies under population inversion conditions has been predicted by Ryzhii et. al. in ~\cite{ryzhii3,ryzhii4} and the potential for graphene based terahertz oscillators was also suggested there. It has been shown that the frequencies of the charge density waves (or plasmons) in graphene at moderate carrier densities (~$10^{9}-10^{12}$ cm$^{-2}$) are in the terahertz frequency range~\cite{ryzhii1,ryzhii2,rana}. Plasmons in graphene are strongly coupled to the interband electronic transitions. The zero bandgap of graphene leads to strong damping of the plasmons at finite temperatures as plasmons can decay by exciting interband electron-hole pairs~\cite{chakra,darma,rana}. Plasmon emission with the accompanying decay of electron-hole pairs has also been experimentally observed in graphene~\cite{ohta}. Recently, the author showed that plasmons in graphene can experience very large gain values (exceeding $10^{4}$ cm$^{-1}$) at frequencies in the 1-10 terahertz range under moderate population inversion conditions due to the stimulated emission of plasmons~\cite{rana}. This process is depicted in Fig.~\ref{Fig1}. Plasmon gain in graphene is similar to the optical gain resulting from the stimulated emission of photons in III-V semiconductor interband lasers at optical and infrared frequencies. The main difference is that the slow group velocity of plasmons in graphene at terahertz frequencies and the strong confinement of the electromagnetic field associated with the plasmons near the graphene layer results in plasmon gain values that are much larger than the typical gain values in semiconductor interband lasers. 

In this paper, we propose and discuss terahertz oscillators based on plasmon amplification in graphene. We present a transmission line model for plasmon propagation in graphene that includes plasmon dissipation due to intraband scattering and plasmon interband gain due to stimulated emission. Using this model, we discuss terahertz plasmon gain in graphene, present designs for practical terahertz oscillators, and derive the threshold condition for oscillation taking into account both intrinsic losses and losses due to coupling power out of the device. A critical parameter that emerges from this analysis is the characteristic impedance of the plasmon transmission line. The threshold condition is shown to depend on the ratio of the external impedance and the characteristic impedance of the plasmon transmission line. The transmission line model presented here is not just relevant for plasmon oscillators but can also be used to model the high frequency electrical behavior of graphene based devices and interconnects. We use this model to obtain the impedance of a graphene strip under population inversion conditions. The electromagnetic energy of plasmons is confined within very small distances of the graphene layer in the vertical direction. In Section~\ref{secpcw}, we discuss techniques to confine plasmons in the lateral direction as well using sub-micron scale dielectric waveguiding structures in order to minimize propagation losses. The large gain values predicted for plasmons in graphene can enable very compact integrated terahertz amplifiers and oscillators. 

Plasmons in metals and semiconductors from THz to optical frequencies have been the subject of much attention in the last few years~\cite{dereux,ozbay}. Plasmon losses in metals are generally large in the visible and near-IR frequency range~\cite{dereux,ozbay}. The strong confinement of the plasmon field in the vicinity of the metal surface near the plasmon resonance frequency makes amplification of plasmons difficult to achieve with a gain medium placed outside the metal. Plasmons in graphene are particularly interesting and unique since the material that supports the plasmons can also provide plasmon gain. The strong confinement of the plasmon field in the vicinity of the graphene layer then becomes an advantage and not an obstacle for achieving plasmon amplification.

\section{Plasmons in Graphene: The Random Phase Approximation (RPA)} \label{secpp}
In this Section, we summarize one of the methods that has been used previously to describe plasmons in graphene and is based on the random phase approximation~\cite{chakra,darma,rana}. In the next Section, we will present a transmission line model for plasmons in graphene that is more relevant for device applications. In graphene, the valence and conduction bands resulting from the mixing of the $p_{z}$-orbitals are degenerate at the inequivalent $K$ and $K'$ points of the Brillouin zone~\cite{dressel}. Near these points, the conduction and valence band dispersion relations can be written as~\cite{dressel},
\begin{equation}
E_{c}({\bf k}) = + \hbar v |{\bf k}| \,\,\,\,\,\,\,\,\,\,\,\,\,\, E_{v}({\bf k}) = - \hbar v |{\bf k}|
\end{equation}
where $v$ is the velocity of the electrons and holes and equals $10^{8}$ cm/s~\cite{dressel}. We assume different Fermi levels for conduction and valence electrons to allow for non-equilibrium electron-hole populations. The electron and hole densities are given by the expressions,
\begin{equation}
n = 4  \int \frac{d^{2}{\bf k}}{(2 \pi)^{2}} \, f(E_{c}({\bf k})-E_{fc})
\end{equation}
\begin{equation}
p = 4  \int \frac{d^{2}{\bf k}}{(2 \pi)^{2}} \, \left[1 - f(E_{v}({\bf k})-E_{fv}) \right]
\end{equation}
The factor of 4 in the front accounts for spin degeneracy and the two valleys at $K$ and $K'$~\cite{dressel}, and $f(E_{c}({\bf k})-E_{fc})$ and  $f(E_{v}({\bf k})-E_{fv})$ are the Fermi distribution functions of the conduction and valence electrons with Fermi energies $E_{fc}$ and $E_{fv}$, respectively. The wavevector ${\bf k}$ is measured from the $K$($K'$) point. The complex propagation vector ${\bf q}(\omega)$ of the plasmon mode of frequency $\omega$ is given by the equation, $\epsilon({\bf q},\omega) = 0$, where $\epsilon({\bf q},\omega)$ is the longitudinal dielectric function of graphene~\cite{chakra,darma,rana}. In this plasmon mode, the charge density (not the number density) associated with the electrons and the charge density associated with the holes oscillate in-phase. In the random phase approximation (RPA) $\epsilon({\bf q},\omega)$ can be written as~\cite{huag},
\begin{equation}
\epsilon({\bf q},\omega) = 1 - V({\bf q}) \Pi({\bf q},\omega) \label{rpa}
\end{equation}
Here, $V({\bf q})$ is the bare 2D Coulomb interaction and equals $e^{2}/2\epsilon_{\infty}q$. $\epsilon_{\infty}$ is the average of the dielectric constant of the media on either side of the graphene layer. $\Pi({\bf q},\omega)$ is the electron-hole propagator including both intraband and interband processes. For small wavevectors (i.e. for $v \, |{\bf q}| < \omega$) the intraband and interband contributions to the propagator can be approximated as follows~\cite{rana},
\begin{IEEEeqnarray}{rcl}
 \Pi_{intra}({\bf q},\omega) & \approx & \frac{q^{2} \, K \, T\,/\pi \hbar^{2}}{ \omega (\omega + i/\tau) - v^{2} \, q^{2}/2} \nonumber \\
& & \times \log \left[ \left( e^{E_{fc}/KT} + 1 \right) \left( e^{-E_{fv}/KT} + 1 \right) \right] \label{propintra}
\end{IEEEeqnarray}
\begin{IEEEeqnarray}{rcl}
\Pi_{inter}({\bf q},\omega) & \approx & \frac{q^{2}}{\hbar} \int_{0}^{\infty} \frac{d{\overline \omega}}{2 \pi} \,  \frac{\left[ f(\hbar {\overline \omega}/2 - E_{fc}) -  f(-\hbar {\overline \omega}/2 - E_{fv}) \right]}{{\overline \omega}^{2} - \omega^{2}} \nonumber \\
& & + i \frac{q^{2}}{4 \hbar \omega} \left[ f(\hbar \omega/2 - E_{fc}) - f(-\hbar \omega/2 - E_{fv}) \right] \label{propinter}
\end{IEEEeqnarray} 
Here, $q=|{\bf q}|$. In Equation (\ref{propintra}), the intraband contribution to the propagator is written in the plasmon-pole approximation that satisfies the f-sum rule~\cite{huag}. This approximation is not valid for large values of the wavevector $q$ when $\omega(q) \rightarrow vq$. However, in this paper we will be concerned with small values of the wavevector for which the plasmons have net gain. Plasmon energy loss due to intraband scattering has been included with a scattering time $\tau$ in the number-conserving relaxation-time approximation which assumes that as a result of scattering the carrier distribution relaxes to the local equilibrium distribution~\cite{mermin}. The real part of the interband contribution to the propagator modifies the effective dielectric constant and leads to a modification in the plasmon frequency under population inversion conditions. The imaginary part of the interband contribution to the propagator incorporates plasmon loss or gain due to stimulated interband transitions. The real and imaginary parts of the propagator in Equations (\ref{propintra}) and (\ref{propinter}) satisfy the Kramers-Kronig relations. Equations (\ref{propintra}) and (\ref{propinter}) can be used with Equation (\ref{rpa}) to calculate the real and imaginary parts of the plasmon frequency $\omega(q)$ as a function of the wavevector $q$. However, from the point of view of device design, it is more useful to assume that the frequency $\omega$ is real and the propagation vector $q(\omega)$, written as a function of $\omega$, is complex.  Equations (\ref{propintra}), (\ref{propinter}), and (\ref{rpa}) can be solved for the complex wavevector $q$. Since the charge density wave corresponding to plasmons has the form $e^{i{\bf q}.{\bf r} - i \omega t}$, the imaginary part of the propagation vector corresponds to net gain or loss.

\begin{figure}[tb]
  \begin{center}
   \epsfig{file=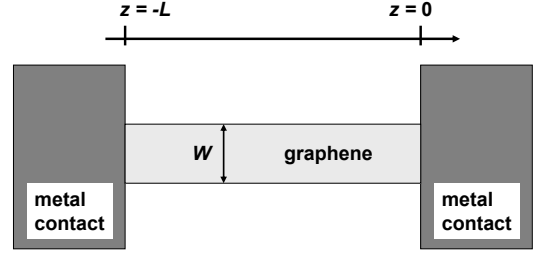,angle=0,width=3.0 in}    
    \caption{A graphene layer connected with metal ohmic contacts on each end (top view). Note that the width $W$ is not the physical width of the graphene layer - but the width of the region in which the plasmons propagate.} 
    \label{Fig2}
  \end{center}
\end{figure}

\section{A Transmission Line Model for Plasmons in Graphene}

\subsection{Transmission LIne Equations}
The procedure for obtaining the plasmon dispersion in graphene outlined in the previous Section does not explicitly account for the connection of the device with the outside world. In this paper, we present a transmission line model for the plasmons in graphene that is more suitable for device design applications. We assume a graphene layer of effective width $W$ and length $L$, connected on each side with a metal ohmic contact to the outside world, as shown in Fig.~\ref{Fig2}. As discussed later in Section~\ref{secpcw}, the width $W$ is not necessarily the physical width of the graphene layer but the effective width of the region in which the plasmons can be confined. We also assume in this Section that population inversion can be achieved in the graphene layer. Details on how to achieve population inversion are discussed in Section~\ref{secpopinv}. We assume a plasmon wave of frequency $\omega$ and wavevector $q$ propagating in the $+z$-direction for which variations in the charge density $Q$ (units: Coulombs/m), current $I_{Q}$ (units: Amps), and the electrostatic potential $V$ (units: Volts) in the plane of the graphene layer can be written as,
\begin{equation}
Q(z,t) = \frac{1}{2} Q(q,\omega)\,e^{i\,q\,z - i\,\omega\,t} + c.c.
\end{equation}       
\begin{equation}
I_{Q}(z,t) =  \frac{1}{2} I_{Q}(q,\omega)\,e^{i\,q\,z - i\,\omega\,t} + c.c.
\end{equation}
\begin{equation}
V(z,t) =  \frac{1}{2} V(q,\omega)\,e^{i\,q\,z - i\,\omega\,t} + c.c.
\end{equation}
Outside the graphene layer the potential decays exponentially as $e^{-q\,|y|}$, where $|y|$ is the distance from the graphene layer. Charge conservation implies,
\begin{IEEEeqnarray}{rcl}
\frac{\partial I_{Q}}{\partial z} & = & - \frac{\partial Q}{\partial t} \nonumber \\
\Rightarrow I_{Q}(q,\omega) & = & \frac{\omega}{q} \, Q(q,\omega) \label{eqcon}
\end{IEEEeqnarray}

\subsection{Electrostatic Capacitance and Quantum Kinetic Inductance} 
The potential $V(q,\omega)$ can be related to the charge density using the electrostatic capacitance per unit length $C_{ES}(q,\omega)$,
\begin{equation}
V(q,\omega) = \frac{Q(q,\omega)}{C_{ES}(q,\omega)} \label{eqcap}
\end{equation}
where $C_{ES}(q,\omega)$ is,
\begin{equation}
C_{ES}(q,\omega) = 2\,\epsilon_{\infty} \,q\,W \label{eqces}
\end{equation}
The expression above for the capacitance assumes a wide ($Wq >> 1$) graphene layer with no top or bottom metal plane. It also assumes that $qL >> 1$. The condition $qL >> 1$ will be assumed to be true throughout this paper. It follows from Equations (\ref{eqcon}) and (\ref{eqcap}) that the current can be related to the potential as follows,
\begin{equation}
I_{Q}(q,\omega) = \frac{\omega \, C_{ES}(q,\omega) }{q}  V(q,\omega)
\end{equation}
A variation in the potential generates a current that can also be expressed using the inductance per unit length $L(q,\omega)$ and the resistance per unit length $R(q,\omega)$ of the plasmon transmission line (as is done in standard transmission line analysis~\cite{kong}),
\begin{equation}
 I_{Q}(q,\omega) = \frac{-i \, q \, V(q,\omega)}{-i\,\omega\,L(q,\omega) + R(q,\omega)} \label{eqI1}
\end{equation}
The inductance $L(q,\omega)$ is the sum of the kinetic inductance $L_{K}(q,\omega)$ and the magnetostatic inductance $L_{MS}(q,\omega)$~\cite{burke1}, and the resistance $R(q,\omega)$ includes plasmon dissipation due to intraband momentum relaxation scattering of electrons and holes as well as plasmon gain due to interband stimulated emission. For all geometries considered in this paper, the magnetostatic inductance is found to be much smaller than the kinetic inductance, and will therefore be ignored. This has been shown to be true for carbon nanotubes as well~\cite{burke1}. The expressions for $L_{K}(q,\omega)$ and $R(q,\omega)$ can be obtained by looking at the microscopic dynamic response of graphene to potential perturbations and this response is described by the electron-hole propagator $\Pi(q,\omega)$~\cite{huag}, 
\begin{equation}
I_{Q}(q,\omega) = \frac{\omega \, e^{2} \, W}{q} \, \Pi(q,\omega) \, V(q,\omega) \label{eqI2}
\end{equation}
The microscopic expressions for $L_{K}(q,\omega)$ and $R(q,\omega)$ can be obtained by comparing the right hand sides of Equations (\ref{eqI1}) and ((\ref{eqI2}). Unlike in the case of carbon nanotubes~\cite{burke1,burke2}, the expression for the kinetic inductance in graphene is frequency and wavevector dependent. The dispersion relation for the plasmons can be obtained from the standard transmission line formula that relates the propagation vector and the frequency~\cite{kong},
\begin{equation}
i\,\omega \, C_{ES}(q,\omega) \,\left[ -i\,\omega\,(L_{K}(q,\omega)+L_{MS}(q,\omega)) + R(q,\omega) \right] = q^{2} \label{eqdisp}
\end{equation}
Assuming $L_{MS}(q,\omega=0$, Equation (\ref{eqdisp}) simplifies to,
\begin{equation}
e^{2} \, W \, \frac{\Pi(q,\omega)}{C_{ES}(q,\omega)} = 1  \label{eqdisp2}
\end{equation}
The above equation for the plasmon dispersion is almost identical to the one obtained by setting $\epsilon(q,\omega)$ equal to zero in the random phase approximation (RPA). The only difference is that Coulomb interaction is expressed in a more general way in terms of the electrostatic capacitance in Equation (\ref{eqdisp2}). Equation (\ref{eqdisp2}), which is obtained using the transmission line model of the plasmons, thus correctly reproduces the plasmon dispersion relation. Given the plasmon dispersion relation, one may drop the explicit functional dependence on the wavevector $q$ in all the expressions. 

\subsection{Power Flow}
The power $P(\omega)$ carried by the plasmon wave can be found using the Poynting vector,
\begin{IEEEeqnarray}{rcl}
P(\omega) & = & \frac{1}{2}\Re \left\{ \int \int dx \, dy \, \, \left[ \vec{E}(\omega) \times \vec{H^{*}}(\omega) \right].\hat{z} \, \right\} \nonumber \\
& = & \frac{1}{2}\Re \left\{\int dy \, \, V(\omega) \, \left[ I_{Q}^{*}(\omega) \, \delta(y) + \frac{\partial \vec{D^{*}}(\omega)}{\partial t} \right] \right\} \label{eqdispnew} \\
& = & \frac{1}{4} \Re \left\{ V(\omega) \,  I_{Q}^{*}(\omega) \right\}  \label{eqdispnew2}
\end{IEEEeqnarray}
The second term in the parenthesis in Equation (\ref{eqdispnew}) is due to the displacement current and arises because unlike in a standard transmission line the plasmon field is not completely transverse to the direction of propagation. The displacement current term's contribution to the power is exactly one-half of the first term and has the opposite sign. Consequently, the expression for the total power in Equation (\ref{eqdispnew2}) is different by a factor of two compared to the standard transmission line formula~\cite{kong}. In order to conserve power when the device is connected to an external circuit, it is convenient to define the transmission line current $I(\omega)$ as,
\begin{equation}
 I(\omega) = \frac{I_{Q}(\omega)}{2} \label{eqcur}
\end{equation}
The current $I(\omega)$ is the maximum current that can be generated by the plasmon wave in an external circuit consistent with the conservation of power. This will be shown explicitly in Section~\ref{secoscillator}. Using this definition of the current, the power carried by the plasmon wave can be written in the standard form,
\begin{equation}
P(\omega) =  \frac{1}{2} \Re \left\{ V(\omega) \,  I^{*}(\omega) \right\}  \label{eqpower}
\end{equation}

\subsection{Plasmon Transmission Line Impedance}
Finally, the expression for the characteristic impedance $Z_{o}(\omega)$ of the plasmon transmission line is defined as,
\begin{IEEEeqnarray}{rcl}
Z_{o}(\omega) & = & \frac{V(\omega)}{I(\omega)} \nonumber \\
& = & 2 \, \sqrt{ \frac{L_{K}(\omega) + i\, R(\omega)/\omega}{C_{ES}(\omega)}} \nonumber \\
& = &  \frac{1}{\epsilon_{\infty} \, \omega \, W}
\end{IEEEeqnarray}

\begin{figure}[tbp]
  \begin{center}
   \epsfig{file=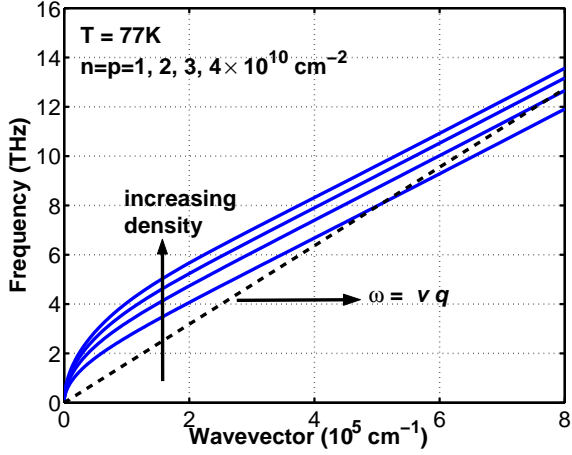,angle=0,width=3.0 in}    
    \caption{Calculated plasmon dispersion relation in graphene at 77K is plotted for different electron-hole densities ($n=p=1, 2, 3, 4 \times 10^{10}$ cm$^{-2}$). The condition $\omega({\bf q}) > v q$ is satisfied for frequencies that have net gain in the terahertz range. The assumed values of $v$ and $\tau$ are $10^{8}$ cm/s and 0.5 ps, respectively.}
    \label{Fig3}
  \end{center}
\end{figure}
\begin{figure}[tbp]
  \begin{center}
   \epsfig{file=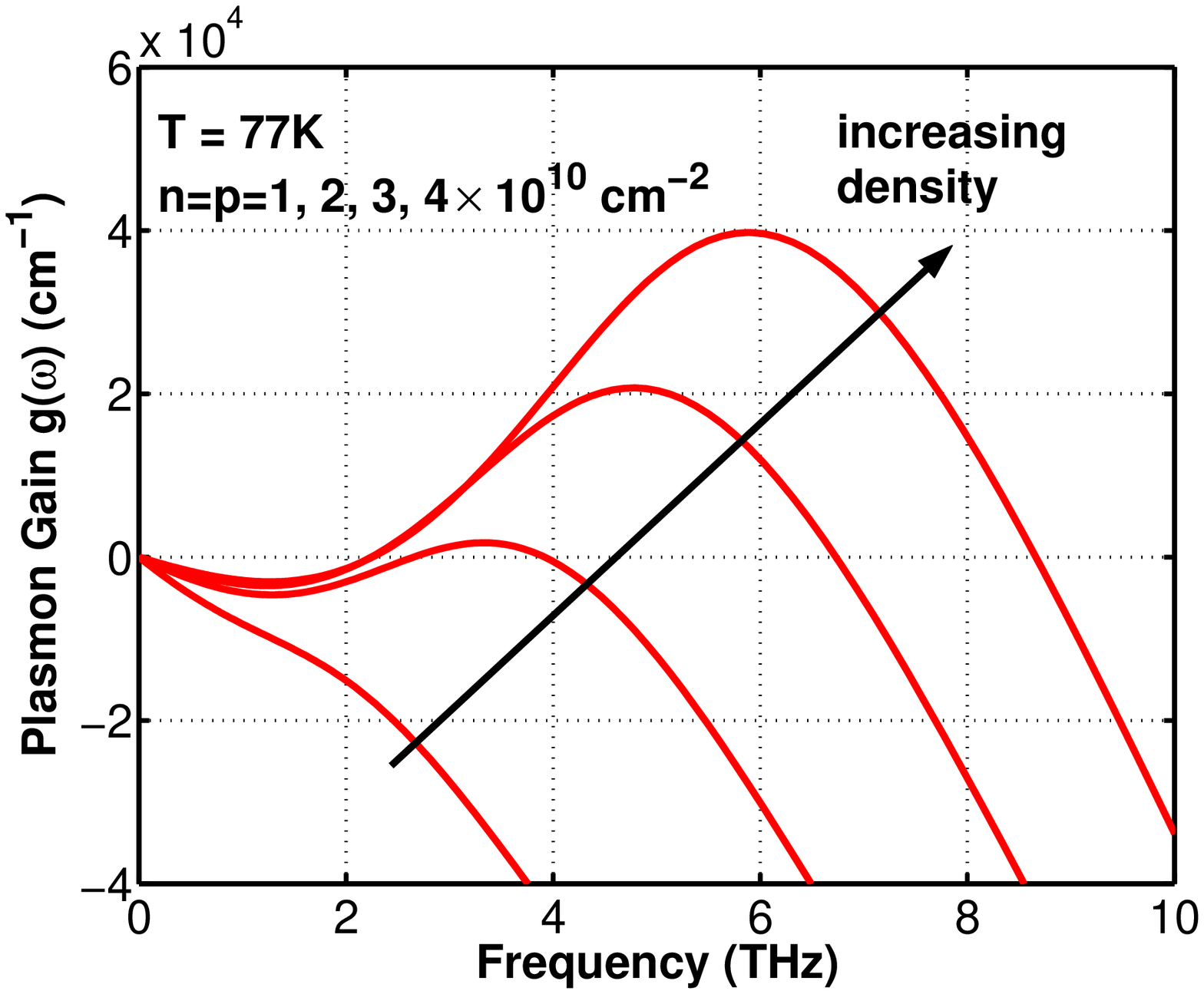,angle=0,width=3.0 in}    
    \caption{Net plasmon gain (interband gain minus intraband loss) in graphene at 77K is plotted for different electron-hole densities ($n=p=1, 2, 3, 4 \times 10^{10}$ cm$^{-2}$). The assumed values of $v$ and $\tau$ are $10^{8}$ cm/s and 0.5 ps, respectively.}  
    \label{Fig4}
  \end{center}
\end{figure}

\begin{figure}[tbp]
  \begin{center}
   \epsfig{file=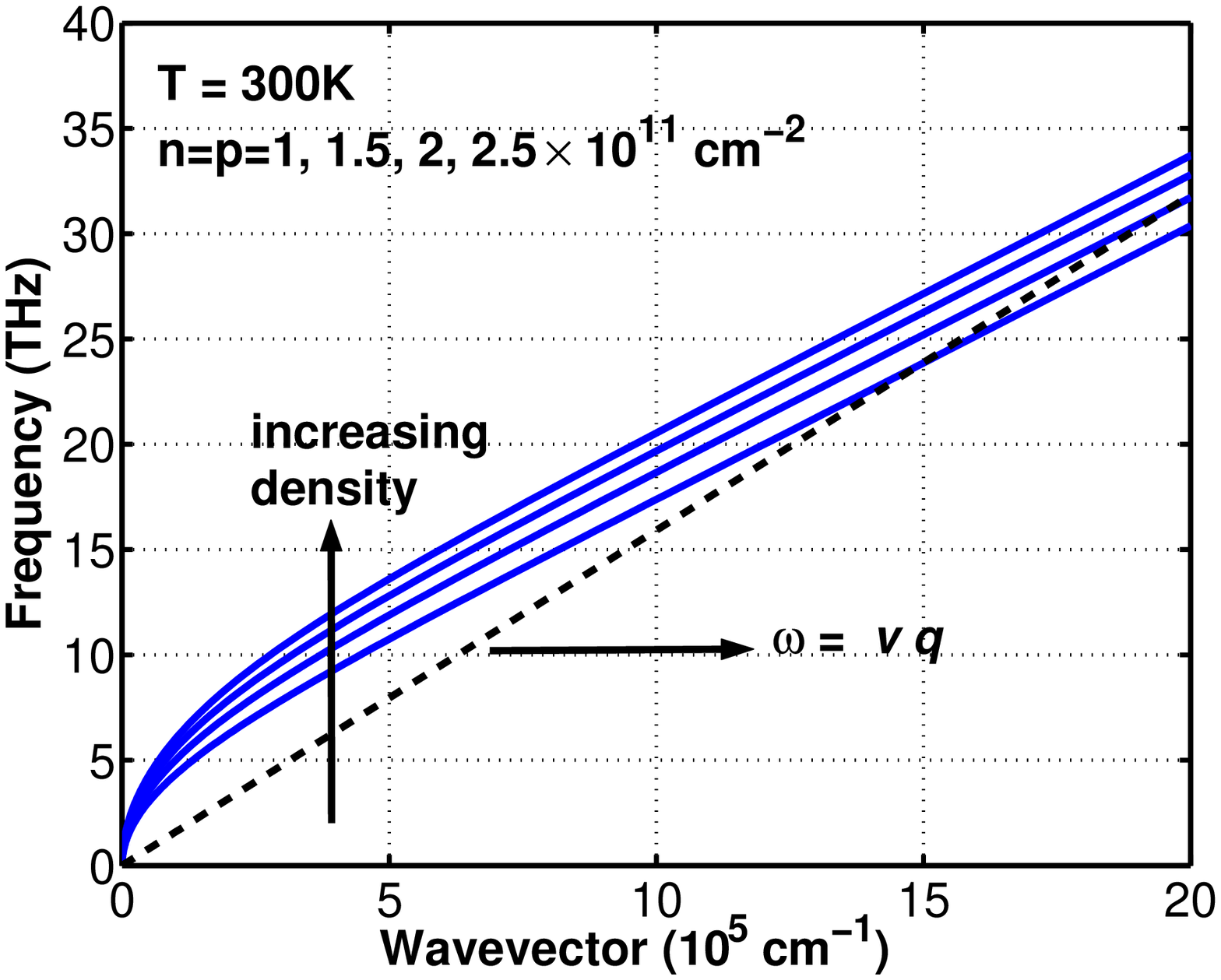,angle=0,width=3.0 in}    
    \caption{Calculated plasmon dispersion relation in graphene at 300K is plotted for different electron-hole densities ($n=p=1, 1.5, 2, 2.5 \times 10^{11}$ cm$^{-2}$). The condition $\omega({\bf q}) > v q$ is satisfied for frequencies that have net gain in the terahertz range. The assumed values of $v$ and $\tau$ are $10^{8}$ cm/s and 0.5 ps, respectively.}
    \label{Fig5}
  \end{center}
\end{figure}

\begin{figure}[tbp]
  \begin{center}
   \epsfig{file=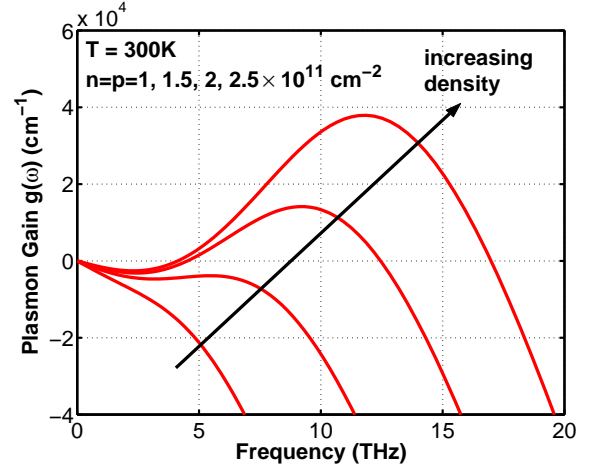,angle=0,width=3.0 in}    
    \caption{Net plasmon gain (interband gain minus intraband loss) in graphene at 300K is plotted for different electron-hole densities ($n=p=1, 1.5, 2, 2.5 \times 10^{11}$ cm$^{-2}$). The assumed values of $v$ and $\tau$ are $10^{8}$ cm/s and 0.5 ps, respectively.}  
    \label{Fig6}
  \end{center}
\end{figure}

\section{Terahertz Plasmon Gain in Graphene}

\begin{figure}[tbp]
  \begin{center}
   \epsfig{file=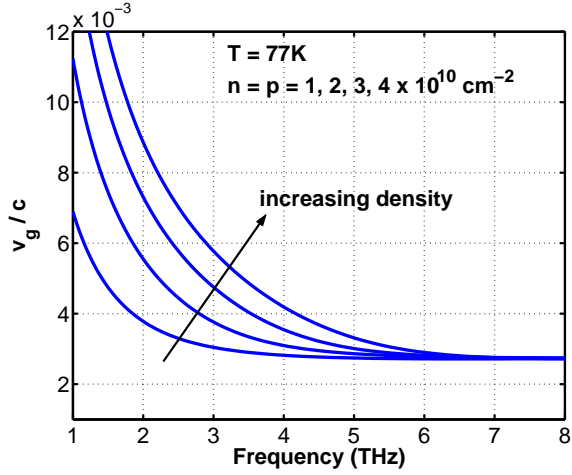,angle=0,width=3.0 in}    
    \caption{The group velocity of the plasmons in graphene, normalized to the speed of light in free space, is plotted for different electron-hole densities ($n=p=1, 2, 3, 4 \times 10^{10}$ cm$^{-2}$) for T=77K. The assumed value of $v$ is $10^{8}$ cm/s.}
    \label{Fig7}
  \end{center}
\end{figure}

In this Section, we discuss the plasmon gain in graphene at terahertz frequencies. Few conditions must be met in order for the plasmons to have net gain. The condition $\Re\{q(\omega)\} < \omega/v$ must be satisfied by the plasmon dispersion relation in order to avoid direct intraband absorption of the plasmons by electrons and holes. The condition $E_{fc} - E_{fv} > \hbar \omega$ must be satisfied in order for the plasmons to have gain via stimulated emission at the frequency $\omega$. In addition, for net gain the loss due to intraband (momentum relaxing) scattering of electrons and holes must be less than the gain due to stimulated interband transitions. This implies that the imaginary part of the propagator $\Pi(q,\omega)$ must be positive (and the resistance $R(q,\omega)$ must be negative). We define the net plasmon energy gain $g(\omega)$ as $-2 \, \Im \{q(\omega)\}$. The net plasmon gain is the difference of the interband plasmon gain and the intraband plasmon losses. Figs.~\ref{Fig3}-\ref{Fig6} show the plasmon dispersion and the net plasmon gain for different electron-hole densities at T=77K and T=300K. Plasmon dispersion (but not plasmon gain) in graphene has been previously reported in ~\cite{chakra, darma,ryzhii1,ryzhii2}. The values used for $v$ and $\epsilon_{\infty}$ in our calculations are $10^{8}$ cm/s and $4\,\epsilon_{o}$, respectively (assuming Silicon-dioxide on both sides of the graphene layer). The calculations assumed an intraband scattering time of 0.5 ps which is less than the experimentally observed scattering time of around 0.6 ps in mobility measurements at T=58K~\cite{rana,heer}. In Ref.~\cite{rana} it has been shown that the plasmon gain remains large even for scattering times as small as 0.1 ps.  The electron and hole densities were assumed to be equal (i.e. $E_{fc} = - E_{fv}$). It can be seen that the net gain approaches $4\times 10^{4}$ cm$^{-1}$ for carrier densities in the $10^{10}$-$5 \times 10^{11}$ cm$^{-2}$ range for frequencies in the 1-10 THz range at T=77K and at T=300K. At low frequencies (less than 1 THz), the losses from intraband scattering tend to dominate. At higher frequencies, the net gain becomes positive for electron-hole densities larger than a minimum threshold value. The calculated plasmon gain values in graphene are between 10 to 100 times larger than the typical material gain values of most III-V semiconductors in the optical and infrared frequency range~\cite{coldren}. This large difference is due to the small group velocity of plasmons in graphene at terahertz frequencies and is similar to the gain enhancement observed in photonic crystals near the photonic band edges due to the small group velocity~\cite{dowling}. The calculated group velocity of plasmons in graphene is shown in Fig.~\ref{Fig7}. The large values for the plasmon gain could enable terahertz amplifiers and oscillators that are only few microns in length. Design of terahertz oscillators and the conditions necessary for achieving terahertz oscillation are discussed in the next Section.

\begin{figure}[tbp]
  \begin{center}
   \epsfig{file=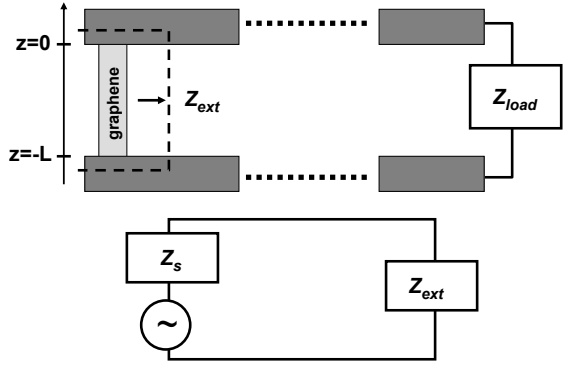,angle=-0,width=3.0 in}    
    \caption{(TOP) A schematic of a graphene terahertz oscillator coupled to an external load. (BOTTOM) Equivalent circuit model.} 
    \label{Fig8}
  \end{center}
\end{figure}

\begin{figure}[tbp]
  \begin{center}
   \epsfig{file=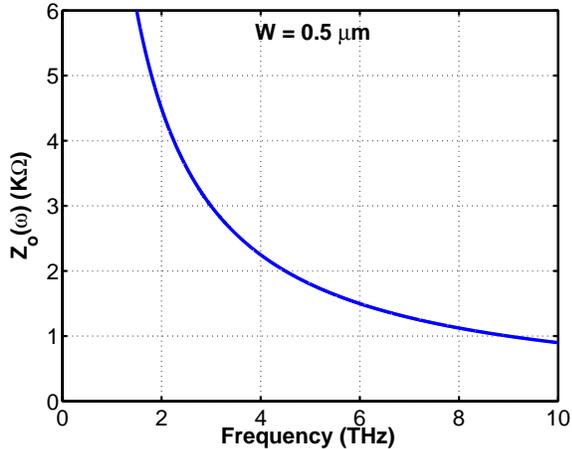,angle=0,width=3.0 in}    
    \caption{The characteristic impedance $Z_{o}(\omega)$ of a $0.5$ $\mu$m wide plasmon transmission line is plotted as a function of the frequency.} 
    \label{Fig9}
  \end{center}
\end{figure}

\begin{figure}[tbp]
  \begin{center}
   \epsfig{file=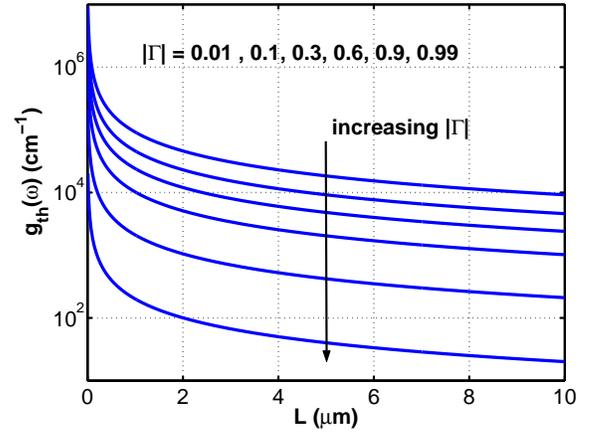,angle=0,width=3.0 in}    
    \caption{The threshold gain $g_{th}(\omega)$ needed for oscillation is plotted as a function of the cavity length $L$ for different values of the parameter $|\Gamma|$.} 
    \label{Fig10}
  \end{center}
\end{figure}

\section{Terahertz Plasmon Oscillators: Design and Oscillation Condition} \label{secoscillator}
We consider the terahertz plasmon oscillator depicted in Fig.~\ref{Fig8} where a population inverted graphene strip of width $W$ and length $L$ is connected to an external circuit. The condition for achieving oscillation in a plasmon oscillator must take into account the energy coupled out of the device since this energy must also be compensated by the plasmon gain in addition to the intrinsic losses that arise due to plasmon loss. We assumed that the external impedance connected to the device (as seen from the two terminals of the device) is $Z_{ext}(\omega)$. $Z_{ext}(\omega)$ also includes the resistances associated with the ohmic contacts to the graphene layer. The external impedance could also represent an antenna structure for coupling the radiation out of the device. We write the current and the potential associated with the plasmons in graphene as a superposition of forward and backward propagating waves using the complex time-harmonic transmission line notation~\cite{kong},
\begin{equation}
V(z) = V_{+}(\omega) \, e^{i\,q\,z} + V_{-}(\omega) \, e^{-i\,q\,z}
\end{equation}
\begin{equation}
I(z) = \frac{V_{+}(\omega)}{Z_{o}(\omega)} \, e^{i\,q\,z} - \frac{V_{-}(\omega)}{Z_{o}(\omega)} \, e^{-i\,q\,z}
\end{equation}  
The reflection coefficient $\Gamma(\omega)$ for the plasmons at the two ends of the graphene strip is defined as,
\begin{equation}
\Gamma(\omega) = \frac{V_{-}(\omega)}{V_{+}(\omega)}
\end{equation}
The boundary conditions imposed by the external circuit are the following,
\begin{equation}
I(z=0) = I(z=-L) \label{eqbc1}
\end{equation}
\begin{equation}
\frac{ V(z=0) - V(z=-L) }{I(z=0)} = Z_{ext}(\omega) \label{eqbc2}
\end{equation}
These boundary conditions give,
\begin{equation}
\Gamma(\omega) = \frac{  Z_{ext}(\omega) - 2\, Z_{o}(\omega)}{ Z_{ext}(\omega) + 2\, Z_{o}(\omega)}  \label{eqgamma}
\end{equation}
and 
\begin{equation}
\Gamma(\omega) \, e^{i\,q\,L} = -1  \label{eqth}
\end{equation}
Equation (~\ref{eqth}) gives the condition necessary for achieving terahertz oscillation. Since $q=\Re\{q\} -i\,g/2$, the threshold gain required to achieve oscillation as well as the frequencies (or the wavevectors) of the oscillating plasmon modes can be obtained from Equation (\ref{eqth}). The threshold gain $g_{th}(\omega)$ needed to achieve oscillation can be expressed as,
\begin{equation}
g_{th}(\omega) = \frac{1}{L} \, \log \frac{1}{|\Gamma(\omega)|^{2}} \label{eqgth}
\end{equation}

The oscillation condition in Equation (\ref{eqth}) can also be obtained using a slightly different approach. The frequency dependent impedance $Z_{s}(\omega)$ of the graphene device looking in from its two terminals at $z=0$ and $z=L$ can also be obtained using the transmission line model and is found to be,
\begin{equation}
Z_{s}(\omega) = 2 Z_{o}(\omega) \frac{ 1 - e^{i\,q\,L} }{1 + e^{i\,q\,L}} \label{eqzs}
\end{equation}
The oscillation condition in Equation (\ref{eqth}) can then be written in the more familiar form as $Z_{s}(\omega) = -Z_{ext}(\omega)$. 

Finally, it needs to be shown that the expression for current in Equation (~\ref{eqcur}) and the boundary conditions in Equations (\ref{eqbc1})-(\ref{eqbc2}) are consistent with the conservation of power between the oscillator and the external circuit. The net power leaving the plasmon oscillator from the two ends at $z=0$ and $z=-L$ can be found using the expression in Equation (\ref{eqpower}),
\begin{equation}
2 \times \frac{|V_{+}(\omega)|^{2}}{2 \, Z_{o}(\omega)} \, \left( 1 - |\Gamma(\omega)|^{2} \right)
\end{equation}
The net power delivered to the load is,
\begin{equation}
\frac{|I(z=0)|^{2}}{2} \Re \left\{ Z_{ext}(\omega) \right\}
\end{equation}
Using Equation (\ref{eqgamma}), the above two expressions can be shown to be equal.

\subsection{Discussion}
The ratio of the characteristic impedance of the plasmon transmission line to the external impedance determines the threshold gain. If the external impedance is much smaller or much greater than the impedance of the transmission line then the threshold gain required for oscillation is close to zero (since $\Gamma \approx \pm 1$). This means that oscillation can be achieved if the gain from stimulated emission overcomes only the internal transmission line losses but no power is coupled out of the device and the device has therefore little practical utility. If in some frequency range $Z_{ext}(\omega) \approx 2\,Z_{o}(q,\omega)$, then $\Gamma(\omega) \approx 0$, and the threshold gain is infinite. In this case, plasmons are not well confined in the cavity and oscillation cannot be achieved for this frequency range. Both situations - when the impedance of the plasmon transmission line is completely mismatched or perfectly matched to the external impedance - need to be avoided. A large capacitive external impedance can also be harmful as it can short the device at high frequencies and result in poor power output coupling efficiencies. Achieving oscillation as well as decent power output coupling efficiencies both therefore depend on the ratio of the external impedance and the characteristic impedance of the plasmon transmission line. The impedance of the plasmon transmission line can range from a few hundred Ohms to a few Kilo-Ohms. Fig.~\ref{Fig9} shows the characteristic impedance $Z_{o}(\omega)$ of a $0.5$ $\mu$m wide plasmon transmission line as a function of the frequency for different carrier densities at T=77K. 

Fig.~\ref{Fig10} shows the threshold gain values needed to achieve oscillation as a function of the cavity length $L$ for different values of the parameter $|\Gamma|$. Comparing the curves in Fig.~\ref{Fig10} to the calculated gain values in Fig.~\ref{Fig4} it can be seen that the large plasmon gain values in graphene could allow plasmon oscillators with cavity lengths much smaller than 10 $\mu$m. As a practical design example, we consider a $0.5$ $\mu$m wide, 5 $\mu$m long, plasmon transmission line connected to an external impedance $Z_{ext}(\omega)$ of $1.0$ K$\Omega$ at 4.0 THz. This value of external impedance would correspond to the impedance of an integrated dipole antenna plus the impedance of the ohmic contacts~\cite{antenna}. We assume $T=77K$ and $\epsilon_{\infty}=4\,\epsilon_{o}$. From Fig.~\ref{Fig9}, the impedance $Z_{o}(\omega)$ of the transmission line at 4.0 THz is around 2.2 K$\Omega$. Therefore, $\Gamma(\omega)$ is -0.63. Using Equation (\ref{eqgth}), the required threshold gain comes out to be  approximately 1800 cm$^{-1}$. This threshold gain value is much smaller than the theoretically predicted maximum net gain values of around 20,000 cm$^{-1}$ available at 4.0 THz in graphene at moderate electron-hole densities (see Fig.~\ref{Fig4}) providing a good margin against extra losses that might not have been accounted for in this paper.

\begin{figure}[tbp]
  \begin{center}
   \epsfig{file=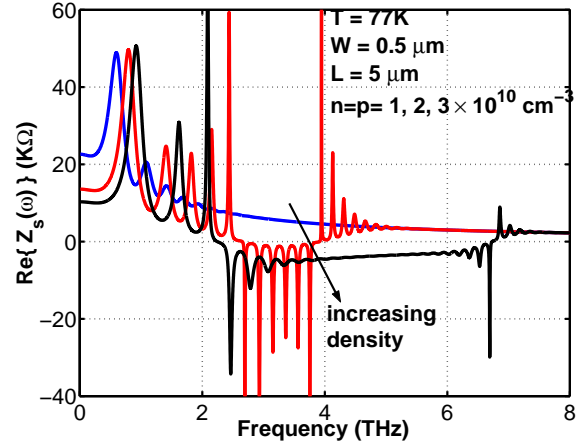,angle=0,width=3.0 in}    
    \caption{The real part of the impedance $Z_{s}(\omega)$ of a $0.5$ $\mu$m wide and  $5$ $\mu$m long graphene strip is plotted as a function of the frequency for different carrier densities at T=77K. The assumed values of $v$ and $\tau$ are $10^{8}$ cm/s and 0.5 ps, respectively.  In the presence of net plasmon gain (e.g., for frequencies between 2.5 and 6.5 THz for $n=p=3 \times 10^{10}$ cm$^{-2}$ - see Fig.~\ref{Fig4}), the real part of the impedance $Z_{s}(\omega)$ is negative.} 
    \label{Fig11}
  \end{center}
\end{figure}

\begin{figure}[tbp]
  \begin{center}
   \epsfig{file=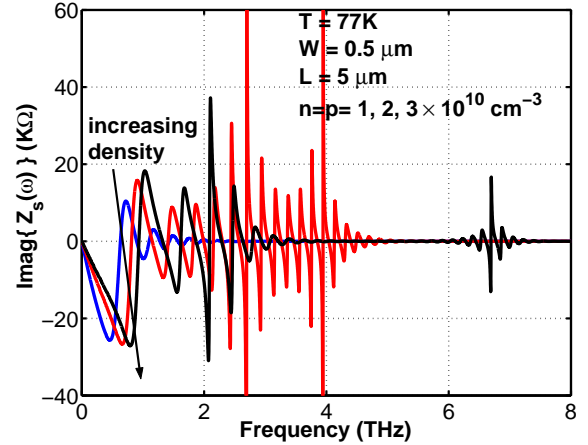,angle=0,width=3.0 in}    
    \caption{The imaginary part of the impedance $Z_{s}(\omega)$ of a $0.5$ $\mu$m wide and  $5$ $\mu$m long graphene strip is plotted as a function of the frequency for different carrier densities at T=77K. The assumed values of $v$ and $\tau$ are $10^{8}$ cm/s and 0.5 ps, respectively.}
    \label{Fig12}
  \end{center}
\end{figure}

The impedance $Z_{s}(\omega)$ of a graphene strip, given in Equation (\ref{eqzs}), is also interesting since in principle it can be experimentally measured at even high frequencies using terahertz time-domain spectroscopy techniques~\cite{thzbook}. The expression in Equation (\ref{eqzs}) indicates that the impedance will exhibit resonances in frequency with spacing given by the inverse plasmon roundtrip time in the graphene strip. These resonances correspond to the confined plasmon modes of the graphene strip. Fig.~\ref{Fig11} and Fig.~\ref{Fig12} show the real and the imaginary parts of the impedance $Z_{s}(\omega)$ of a $0.5$ $\mu$m wide and  $5$ $\mu$m long graphene strip as a function of the frequency for different electron and hole densities at T=77K. The plasmon resonances in the impedance are clearly visible. The resonances disappear when the plasmon loss or the plasmon gain are large, and become pronounced when the net plasmon gain is close to zero. In the presence of net plasmon gain (e.g., for frequencies between 2.5 and 6.5 THz in Fig.~\ref{Fig11} and Fig.~\ref{Fig12} for $n=p=3 \times 10^{10}$ cm$^{-2}$ - see Fig.~\ref{Fig4}), the real part of the impedance  $Z_{s}(\omega)$ is negative. Experimental measurement of the impedance $Z_{s}(\omega)$ under population inversion conditions can therefore be used to measure the plasmon gain in graphene. Note that at very small frequencies, when $qL >> 1$ does not hold, the expression for the impedance in Equation (\ref{eqzs}) is not valid.  

\begin{figure}[tbp]
  \begin{center}
   \epsfig{file=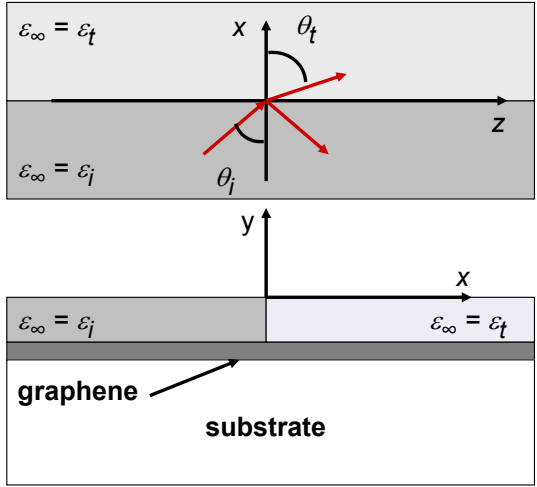,angle=0,width=3.0 in}    
    \caption{Plasmon propagation across a dielectric interface in graphene: (TOP) Top view (BOTTOM) Side view.}  
    \label{Fig13}
  \end{center}
\end{figure}

\section{Plasmon Confinement and Waveguiding} \label{secpcw}
\begin{figure}[tbp]
  \begin{center}
   \epsfig{file=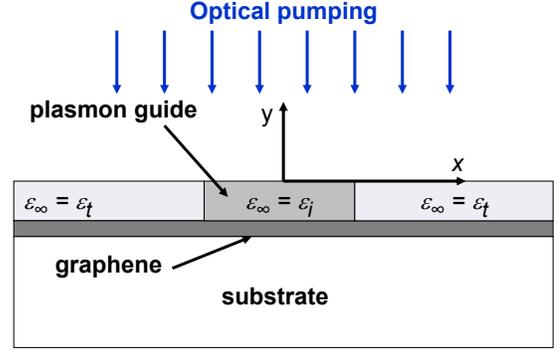,angle=0,width=3.0 in}    
    \caption{A cross-sectional view of a hybrid waveguide-transmission line structure for plasmon confinement and guiding. Population inversion is achieved by optical pumping.} 
    \label{Fig14}
  \end{center}
\end{figure}

In Section~\ref{secpp}, we mentioned that the width $W$ is not necessarily the physical with of the graphene layer. In this Section, we show that plasmons can be confined laterally and guided in hybrid waveguide-transmission-line structures. This is useful since plasmons need to be kept away from regions where population inversion is small in order to minimize losses. To illustrate the basic idea we consider a graphene layer on top of an insulating substrate. The top surface of the graphene layer is covered by two insulating media of different dielectric constants such that the value of $\epsilon_{\infty}$ in the two cases are $\epsilon_{i}$ and $\epsilon_{t}$, and these two media form an interface at x=0, as shown in Fig.~\ref{Fig13}. Consider a plasmon wave incident at an angle $\theta_{i}$ at the interface. The transmitted wave makes an angle $\theta_{t}$. The condition for phase matching~\cite{kong}, and the continuity of the potential and the normal component of the current at the interface can give the value of the angle $\theta_{t}$ and the amplitudes of the reflected and the transmitted plasmon waves. The phase matching condition gives, 
\begin{equation}
q_{i}(\omega) \,\, \sin(\theta_{i}) = q_{t}(\omega) \,\, \sin(\theta_{t})
\end{equation}
For small wavevectors, the above Equation can be approximated with very little error by, 
\begin{equation}
\epsilon_{i} \,\, \sin(\theta_{i}) = \epsilon_{t} \,\, \sin(\theta_{t})
\end{equation}
If  $\epsilon_{i} > \epsilon_{t}$, then for $\theta_{i}$ larger than a certain critical angle $\theta_{c}$ the plasmon wave is totally internally reflected. The value of the critical angle  $\theta_{c}$ is $\sin^{-1}( \epsilon_{t}/ \epsilon_{i})$. For $\theta_{i}$ larger than the certain critical angle, the incident plasmon wave is completely reflected. This suggests that it is possible to confine and guide plasmons in hybrid waveguide-transmission line structures in which vertical confinement is provided by the graphene layer itself and lateral confinement is provided by the dielectric guide that has a single core sandwiched between two claddings on either side. A cross-sectional view of such a guide is shown in Fig.~\ref{Fig14}. The plasmon guide, like dielectric optical waveguides~\cite{kong}, can support more than one transverse mode. A single transverse mode plasmon guide is desirable for efficient power coupling to the external circuit and can be obtained by keeping the difference between the core and cladding $\epsilon_{\infty}$ values small. A full analysis of the modes supported by the guide shown in Fig.~\ref{Fig14} is beyond the scope of this paper but we should mention here that for a single transverse mode plasmon guide the width $W$ used in the analysis is roughly equal to the effective width of the transverse mode in the guide. Methods to achieve population inversion in graphene are discussed in the next Section.

\begin{figure}[tbp]
  \begin{center}
   \epsfig{file=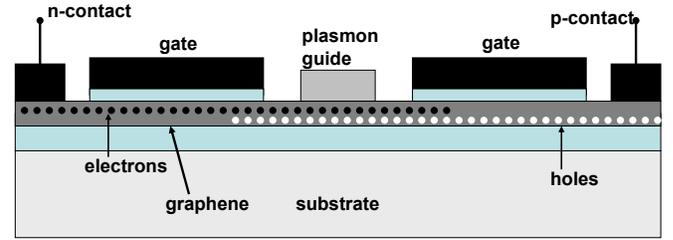,angle=0,width=3.5 in}    
    \caption{A cross-sectional view of a plasmon guide integrated on a gated pn-junction in graphene. Population inversion is achieved via electron-hole injection under an applied forward bias.}
    \label{Fig15}
  \end{center}
\end{figure}

\section{Achieving Population Inversion in Graphene} \label{secpopinv}
Schemes for achieving population inversion in graphene via optical and electrical pumping have been extensively discussed by Ryzhii et. al. in~\cite{ryzhii3,ryzhii4} in the context of obtaining negative conductivity in graphene at terahertz frequencies. A knowledge of the electron-hole recombination times in graphene is critical for the evaluation of different schemes but no experimental results have so far been reported on the recombination times in graphene. Small band-gap semiconductors usually have large electron-hole recombination rates due to Coulomb scattering (Auger recombination)~\cite{paul}. The zero bandgap and the large optical phonon energy in graphene ($>$ 180 meV~\cite{ando}) suggest that electron-hole recombination rates could be dominated by Auger processes. Results in Ref.~\cite{rana2} show that electron-hole recombination times due to Auger scattering in graphene are weakly temperature dependent and can range from 1 ps to 100 ps for electron-hole densities in the $10^{9}$-$10^{12}$ cm$^{-2}$ range. Optical pumping scheme for achieving population inversion in a plasmon waveguide is depicted in Fig.~\ref{Fig14}. The electron-hole generation rate $G_{P}$ in graphene under optical pumping with pump intensity $I_{P}$ and pump frequency $\omega_{P}$ is given by the expression~\cite{ryzhii4,dressel2},
\begin{equation}
G_{P} \approx \frac{e^{2}}{4 \, \hbar \, \sqrt{ \epsilon_{\infty} \, \epsilon_{o}} \, c} \, \frac{I_{p}}{\hbar \omega_{P}}
\end{equation}
Assuming T=77K, an Auger recombination time of around 10 ps for electron-hole densities in the low $10^{10}$ cm$^{-2}$ range (see Ref.~\cite{rana2}), and optical pumping with a 10 $\mu$m wavelength laser, achieving plasmon gain at around 4.0 THz would require a pump intensity of 2 KW/cm$^{2}$. This intensity level is easily achievable with a focused beam from a standard 5-Watt CO$_{2}$ laser. High power Q-switched pump lasers can also be used at shorter near-IR wavelengths for pulsed operation. Another way to achieve population inversion in graphene is via electron-hole injection in a forward biased electrostatically gated pn-junction, as is done in semiconductor interband lasers~\cite{marcus,ryzhii3}. This scheme is depicted in Fig.~\ref{Fig15} which shows the cross-section of a plasmon guide at the junction of electrostatically gated p and n regions. The guide helps to confine the plasmons in the region where population inversion is expected to be maximum under an applied forward bias. The zero-bandgap of graphene suggests that band-to-band tunneling current could be large in a forward biased pn-junction and could possibly limit electron-hole injection. In Ref.~\cite{ryzhii3} it was shown that the injected components of the electron and hole currents in a forward biased graphene pn-junction could be significant and could allow for population inversion to be achieved in the vicinity of the junction.     

\section{Challenges and Conclusion}
In this paper, we proposed and analyzed terahertz oscillator designs based on plasmon amplification in graphene via stimulated emission. We presented a transmission line model for plasmon propagation and amplification in graphene, and obtained the condition for oscillation that took into account plasmon losses due to intraband scattering and also losses due to external coupling. The ratio of the characteristic impedance of the plasmon transmission line to the external impedance was shown to play an important role in determining the threshold condition for oscillation. The large values of the plasmon gain could allow the realization of compact integrated terahertz oscillators. 

Several obstacles need to be overcome before the proposed oscillators can be realized. Material quality remains a challenge. Graphene monolayers and multilayers produced from currently available experimental techniques are estimated to have defect/impurity densities anywhere between $10^{11}$ and $10^{12}$ cm$^{-2}$ and single-crystal coherence lengths of less than a micron~\cite{heer,darmatr}. The values of the electron and hole momentum scattering times are critical in determining plasmon losses. We modeled the intraband scattering rate with a single time constant $\tau$ the value of which was estimated from experimental measurements of carrier mobility in graphene. Several momentum scattering mechanisms that could contribute to plasmon losses might have not been accounted for in this approach. For example, interband momentum relaxing scattering via acoustic phonons could contribute to plasmon losses under population inversion conditions for a zero-bandgap material. The carrier density dependence of the scattering time was also ignored. Plasmon losses due to image currents flowing in surrounding metals, ground plane, etc, were not included in the analysis presented here. Experimental investigations will be needed to provide answers to many of these questions. Given that plasmon wavelengths at terahertz frequencies are much shorter than one micron (see Fig~\ref{Fig3}), surface roughness of the plasmon guide could scatter plasmons and also contribute to plasmon losses. It is also not clear at the time of the submission of this manuscript if or not graphene has a small bandgap. Several factors, such as disorder, substrate interaction, and spin-orbit interaction, can lead to the opening of a small bandgap in graphene. The main ideas presented in this paper are expected to remain valid qualitatively provided the bandgap is smaller than or equal to the frequency of the plasmons.    

The maximum output power levels achievable with the proposed oscillators would be determined by the maximum rate at which electrons and holes can be injected, device heating, or nonlinear effects (such as band-to-band tunneling caused by the plasmon electric fields). Nonlinear effects would also become important because of the simple fact that the amplitude of the charge density wave can never exceed the actual charge density of the electrons and the holes in the device. In fact, assumptions made in this paper will break down well before the amplitude of the charge density wave reaches such levels. A rough estimate of the power levels can be made as follows. Assuming an electron-hole density of $6\times10^{10}$ cm$^{-2}$, frequency of oscillation of 4 THz, T=77K, a plasmon guide of width $0.5$ $\mu$m and length $5$ $\mu$m, the values of $q(\omega)$ and $Z_{o}(\omega)$ come out to be approximately $10^{5}$ cm$^{-1}$ and 2250 $\Omega$, respectively. The separation of the electron and hole Fermi levels is approximately 50 meV. Assuming a density wave amplitude of $2\times 10^{10}$ cm$^{-2}$, the amplitude of the potential associated with the wave comes out to be approximately 45 mV. This gives a power of the order of 1 $\mu$Watt. The compact dimensions of the proposed oscillators could allow small phase-locked arrays to increase the output power to mWatt levels. 
 
The author would like to thank Edwin Kan, Sandip Tiwari, Michael Spencer, Faisal R. Ahmad, and Hassan Raza for helpful discussions.

\newpage

\begin{IEEEbiography}{Farhan Rana}
Farhan Rana obtained BS, MS, and PhD degrees all in Electrical Enginnering from the Massachusetts Institute of Technology (MIT), Cambridge, MA (USA). He worked on a variety of different topics related to semiconductor optoelectronics, quantum optics, and mesoscopic physics during his PhD research. Before starting PhD, he worked at IBM's T. J. Watson Research Center on Silicon nanocrystal and quantum dot memory devices. He finished PhD in 2003 and joined the faculty of Electrical and Computer Engineering at Cornell University, NY. His current researh focuses on semiconductor optoelctronics and terahertz devices. He is the recepient of the US National Science Foundation Faculty CAREER award in 2004.      
\end{IEEEbiography}

\end{document}